\documentstyle[preprint,eqsecnum,aps]{revtex}
\tighten
\begin{document}
\preprint{Alberta-Thy-37-96}
\draft

\title{Are Higher Order Membranes stable in Black Hole Spacetimes?}
\author{A. L. LARSEN\thanks{Electronic Address:
alarsen@phys.ualberta.ca}}
\address{Theoretical Physics Institute, Department of Physics,\\
University of Alberta, Edmonton, Canada T6G 2J1}
\author{C. O. LOUSTO\thanks{Electronic Address:
lousto@mail.physics.utah.edu}}
\address{Department of Physics, University of Utah,\\
201 JFB, Salt Lake City, UT 84112, USA}
\date{\today}
\maketitle
\begin{abstract}

We continue the study of the existence and stability of static spherical
membrane configurations in curved spacetimes. We first consider
higher order membranes described by a Lagrangian which, besides the Dirac
term, includes a term proportional to the scalar curvature
of the world--volume ${}^{(3)}R$. Notably, in this
case, the equations of motion can be reduced to second order ones
and an effective potential analysis can be made. The conditions
for stability are then explicitly derived. We
find a self--consistent static spherical membrane, determining the spacetime
generated by the membrane itself. In this
case we find, however, that the total energy
of the membrane has to be negative, and no {\it stable} equilibrium can
be achieved.
We then generalize the discussion to a membrane described by a Lagrangian
including all possible second derivative terms.
We conclude the paper with some discussion on the
generality of the results obtained.

\end{abstract}

\pacs{11.27.+d,04.70.Dy}

\section{Introduction}

In Ref.\ \cite{LL96} we have started the investigation of the plausibility
of having a stable spherical membrane in the curved background of
a (spherically symmetric) black hole, with regards to the possibility
of the membrane being able to represent, in an effective way, the quantum
degrees of freedom of the event horizon\ \cite{L95}.
We have found that, contrary
to what happens in flat spacetime for bosonic Dirac
membranes\ \cite{D62}, static equilibrium solutions are
possible. In fact, as an example we have
given explicitly the equilibrium radii
$r_m$ for the Schwarzschild--de Sitter background metric. Although
equilibrium is reached, it is not stable against perturbations. We
have identified the mode $l=0$ (and under certain conditions also
$l=1$) as the responsible for the instability. We have then considered
higher order membranes \cite{GG90,CG95,BGMPU95} (coupling to the extrinsic
curvature), which
can be thought of as finite thickness membranes, as opposite to Dirac
membranes which have zero thickness. Also in this case it is known that
no static spherical
equilibrium solutions are possible in flat spacetime \cite{C94}. We have
analyzed the stability of the higher order membranes in
the Schwarzschild background, finding that equilibrium solutions actually do
exist there, but they are unstable.
However, this by no means proves in general the non--existence of stable
equilibrium solutions in curved spacetimes.

Two possibilities are now open: One could try to
prove some kind of ``no hair'' theorem for the case of membranes 
coupled only to the gravitational field, i.e., try to prove that it is
actually impossible to have stable equilibrium membranes around a black hole.
A second possibility,
and this is the one we will follow in this paper, is to look at the
back-reaction problem, i.e., to solve self-consistently the equations of
motion representing a membrane plus black hole system in equilibrium,
and try to provide a counter--example to the eventual ``no hair'' theorem for
membranes.

The paper is organized as follows: In Sec.\ II we discuss the higher order
membrane Lagrangian in general terms.
Sec.\ III deals with
the higher order membrane  Lagrangian including only the term  proportional
to the scalar curvature of the world--volume. In this case
an effective potential analysis can be made and the conditions
for stability are then explicitly derived. In Sec.\ IV we
find a self--consistent static spherical membrane coupled
to Einsteinian gravity, that is to say,
we determine the spacetime generated by the membrane instead of
fixing it {\it a priori}. This self-consistent membrane is however also
unstable. In Sec.\ V we then generalize the discussions of  Secs.\ III
and IV to the more complicated case of the
membrane Lagrangian including all the possible
second derivative terms. We finally end the paper with a discussion on
the generality of the unstable behavior of membranes in black
hole backgrounds.

\section{Higher order membrane in a spherically symmetric curved
spacetime}

Up to second derivatives in the membrane world-volume coordinates,
the most general action can be written as\cite{GG90,CG95,BGMPU95}

\begin{equation}
S_{m}=\int d\tau d\rho d\sigma  \sqrt{-
\gamma}
\left[-T+A\left(\gamma^{ij}\Omega_{ij}\right)^2+
B\ \Omega^{ij}\Omega_{ij}+C\ {}^{(3)}R\right]~,
\label{IV.1}\end{equation}
where $\gamma_{ij}$ is the induced metric on the world-volume
\begin{equation}
\gamma_{ij}=g_{\mu\nu} X^{\mu}_{,i}X^{\nu}_{,j}~,
\label{1.1}
\end{equation}
$\gamma$ is its determinant and ${}^{(3)}R$ is its scalar curvature;
$g_{\mu\nu}$ is the spacetime
metric, while
\begin{equation}
\Omega_{ij}=g_{\mu\nu}n^\mu X^\rho_{,i}\nabla_\rho X^\nu_{,j}~,
\label{3.3}\end{equation}
is the second fundamental form (extrinsic curvature), where the
normal vector $n^\mu$ is defined by
\begin{equation}
g_{\mu\nu}n^\mu X^\nu_{,i}=0,\;\;\;\;\;\;\;\;\;\;
g_{\mu\nu}n^\mu n^\nu=1,\label{2.4}
\end{equation}
and it fulfills the completeness relation
\begin{equation}
g^{\mu\nu}=n^\mu n^\nu+\gamma^{ij} X^\mu_{,i} X^\nu_{,j}.
\end{equation}
Notice that the tension $T$ has dimension of $length^{-3}$ while
the arbitrary
constants $A$, $B$, and $C$ carry dimension of $length^{-1}$.

In Ricci--flat spacetimes, the scalar curvature
of the world--volume is related to the two second fundamental
form terms, via the Gau\ss--Codazzi equation
\begin{equation}
{}^{(3)}R=\left(\gamma^{ij}\Omega_{ij}\right)^2-
\Omega^{ij}\Omega_{ij}~,\label{2.6}
\end{equation}
and therefore only two of the last three terms in Eq.\ (\ref{IV.1})
are independent. However, in this paper we will also be interested in
studying non--Ricci--flat spacetimes, where all terms should be
included.

We shall consider static and spherically symmetric backgrounds
\begin{equation}
ds^2=-a(r)dt^2+b(r)^{-1}dr^2+r^2d\Omega^2~,~~
d\Omega^2=d\theta^2+\sin^2\theta d\varphi^2~.\label{1.3}
\end{equation}

A spherical membrane with time-dependent radius can
be conveniently described by the following spherically symmetric
rest gauge choice
\begin{equation}
t=\tau~,~~r=r(\tau)~,~~\theta=\rho~,~~\varphi=\sigma~,
\label{1.3'}
\end{equation}
so that the induced metric on the world--volume becomes
\begin{eqnarray}
&\gamma_{\tau\tau}=-a+\dot r^2/b~~,~~~
\gamma_{\rho\rho}=r^2~,~~
\gamma_{\sigma\sigma}=r^2\sin^2\rho~&,\nonumber\\
& &\\
&\sqrt{-\gamma}=r^2\sin\rho\sqrt{a-\dot r^2/b},~&\label{2.9}\nonumber
\end{eqnarray}
where a dot denotes derivative with respect to $\tau$.

{}From $\gamma_{ij}$ we can now compute the non--vanishing
components of the (world--volume) Ricci tensor
\begin{eqnarray}
{}^{(3)}R_{\tau\tau}&=&{-\dot r^2\over r\left(a-\dot r^2/b\right)}
\left(-a'+2\ddot r b^{-1}-\dot r^2b'b^{-2}
\right)-{2\ddot r\over r}~,\\
{}^{(3)}R_{\rho\rho}&=&1+{\dot r^2+r\ddot r\over\left(a-\dot r^2/b\right)}
+{r \dot r^2\over 2\left(a-\dot r^2/b\right)^2}
\left(-a'+2\ddot r b^{-1}-\dot r^2b'b^{-2}\right)~,\\
{}^{(3)}R_{\sigma\sigma}&=&\sin^2\rho\;\; {}^{(3)}R_{\rho\rho}~,
\end{eqnarray}
where a prime denotes derivative with respect to $r.$
{}From these expressions, it is easy to compute ${}^{(3)}R$
\begin{equation}
{}^{(3)}R=\frac{2a}{r\left(a-\dot r^2/b\right)^2}(2\ddot{r}-
a'b-b'a)+\frac{2(a'b+2b'a)}{r\left(a-\dot
r^2/b\right)}+\frac{2ab}{r^2\left(a-\dot r^2/b\right)}+
\frac{2(1-b)-2rb'}{r^2}.\label{2.13}
\end{equation}

We shall also need the components of the second fundamental form. The normal
vector introduced in Eq.\ (\ref{2.4}) is given by:
\begin{equation}
n^\mu=\frac{\sqrt{b/a}}{\sqrt{a-\dot r^2/b}}\left(\dot{r}/b,\;a,\;0,\;0\right).
\end{equation}
and it is then straightforward to obtain explicit expressions for
$\Omega_{ij}$
\begin{equation}
\Omega_{\tau\tau}=\frac{a\sqrt{b/a}}{2b{\sqrt{a-\dot r^2/b}}}(2\ddot{r}-
a'b-b'a)+\frac{\sqrt{a-\dot r^2/b}}{a\sqrt{b/a}}(a'b+b'a/2),
\end{equation}
\begin{equation}
\Omega_{\rho\rho}=\frac{-ar\sqrt{b/a}}{\sqrt{a-\dot r^2/b}},
\end{equation}
\begin{equation}
\Omega_{\sigma\sigma}=\sin^2\rho\;\Omega_{\rho\rho}
\end{equation}
It follows that
\begin{eqnarray}
(\gamma^{ij}\Omega_{ij})^2&=&\frac{a}{4b(a-\dot r^2/b)^3}
(2\ddot{r}-a'b-b'a)^2+\frac{(2ab+r(a'b+b'a/2))^2}{r^2ab
(a-\dot r^2/b)}\nonumber\\
&+&\frac{(2ab+r(a'b+b'a/2))}{rb(a-\dot r^2/b)^2}(2\ddot{r}-a'b-b'a),
\label{2.18}
\end{eqnarray}
while
\begin{eqnarray}
\Omega_{ij}\Omega^{ij}&=&\frac{a}{4b(a-\dot r^2/b)^3}
(2\ddot{r}-a'b-b'a)^2+\frac{(a'b+b'a/2)^2}{ab
(a-\dot r^2/b)}\nonumber\\
&+&\frac{(a'b+b'a/2)}{b(a-\dot
r^2/b)^2}(2\ddot{r}-a'b-b'a)+\frac{2ab}{r^2(a-\dot r^2/b)}.
\label{2.19}
\end{eqnarray}
Comparing with Eq.\ (\ref{2.13}) we see that
\begin{eqnarray}
{}^{(3)}R-\left(\gamma^{ij}\Omega_{ij}\right)^2+
\Omega^{ij}\Omega_{ij}&=&\frac{2(b'a-a'b)}{r\left(a-\dot
r^2/b\right)}+\frac{2(1-b)-2rb'}{r^2}\nonumber\\
&=&{}^{(4)}R-2\;{}^{(4)}R_{\mu\nu}n^\mu n^\nu,
\end{eqnarray}
in agreement with the Gau\ss--Codazzi equation for non-Ricci-flat
spacetimes of the form\ (\ref{1.3}).
\section{Membrane with Scalar Curvature Term}

For the sake of simplicity we will consider in this section
the reduced model of Eq.\ (\ref{IV.1}) with $A=0=B$. This resulting
action is essentially the 3--dimensional Einstein action with a
cosmological constant. However, here, we must take the functional
variation with respect to $X^\mu(\tau,\sigma,\rho)$ {\em instead}
of $\gamma_{ij}$! This makes the generic equations of motion rather
complicated. In this paper we are though interested in spherical
membranes and, therefore, it is easier to derive the equations of
motion from the effective Lagrangian
\begin{eqnarray}
L&=&-4\pi Tr^2\sqrt{a-\dot r^2/b}\nonumber\\
&+&8\pi C\left\{{ar(2\ddot r-a'b-b'a)\over
\left(a-\dot r^2/b\right)^{3/2}}+{r(a'b+2b'a)+ab
\over\sqrt{a-\dot r^2/b}}+(1-b-rb')\sqrt{a-\dot r^2/b}
\right\}~,
\label{lagrangiano}
\end{eqnarray}
as obtained from equations (\ref{IV.1}),(\ref{2.9}) and (\ref{2.13}).
Note that the Lagrangian depends on $r,\dot r$ and $\ddot r$. The
standard way to deal with such situations is as follows:

We build up the conjugate momenta as
\begin{equation}
P_1\dot={\delta L\over\delta\dot r}-
\partial_\tau \left( {\delta L\over\delta\ddot r}\right)~,~~
 P_2\dot={\delta L\over\delta\ddot r}~.\label{3.2}
\end{equation}
The equations of motion then read
\begin{equation}
\dot{P_1}-{\delta L\over\delta r}=0~\label{3.3'}
\end{equation}
and the Hamiltonian
\begin{equation}
{\cal H}=P_1\dot r+P_2\ddot{r}-L~.\label{3.4}
\end{equation}
Note that this Hamiltonian is conserved in the usual sense:
$\dot{\cal H}=0$, i.e. ${\cal H}=E=constant$ [this can be explicitly
checked by the use of the generalized Euler--Lagrange equations (\ref{3.3'})].

In our case, the Hamiltonian is explicitly given by
\begin{equation}
{\cal H}=4\pi T{ar^2\over\sqrt{a-\dot r^2/b}}-8\pi C\left[
{a(1-b)\over\sqrt{a-\dot r^2/b}}+{a^2b\over
\left(a-\dot r^2/b\right)^{3/2}}\right]~.\label{hamiltoniano}
\end{equation}
Note that the higher derivatives of $r$ have canceled out!
This allows us to define an effective potential to analyze the
stability of the membrane. In fact, ${\cal H}=E$ and
Eq.\ (\ref{hamiltoniano}) leads to
\begin{equation}
E^2\dot r^2=E^2-V(r)^2~,\label{3.6}
\end{equation}
where $E$ has dimension of $length^{-1}=mass$ in units where
$c=\hbar=1$ but $G$ is kept explicitly, and $V(r)$ is the effective potential
that can
be explicitly obtained from \ (\ref{hamiltoniano}) by inversion
of a cubic equation.

The conditions for the existence of a static equilibrium solution at
$r=r_m$, i.e., $\dot r=0,\; V^2=E^2$ and $(V^2)'=0$ for $r=r_m,$ are
\begin{eqnarray}
E&=&4\pi \sqrt{a}\;(Tr^2-2C)\Biggr\vert_{r_m}\label{equilibrio1}\\
{T\over C}&=&{2a'\over(r^2a'+4ar)}\Biggr\vert_{r_m}~.
\label{equilibrio}
\end{eqnarray}
These two equations can be seen as determining the values of
$E$ and $r_m$ for the equilibrium membrane for given values of $T$ and $C$.
Alternatively, for a given value of $r_m,$ equation 
(\ref{equilibrio}) determines the value of
$T/C$ necessary to support such a membrane.

This equilibrium solution will be stable when $(V^2)''>0$ for $r=r_m$, i.e.
\begin{equation}
{T\left[{r^2a''\over2\sqrt{a}}-{r^2a'^2\over4a^{3/2}}
+{2ra'\over\sqrt{a}}+2\sqrt{a}\right]-C\left[{a''\over
\sqrt{a}}-{a'^2\over2a^{3/2}}\right]
\over\left[{Tr^2\over2b\sqrt{a}}-C\left({1\over
b\sqrt{a}}+{2\over\sqrt{a}}\right)\right]}\Biggr\vert_{r_m}>0~.
\label{estabilidad}
\end{equation}
[Note that both the equilibrium condition (\ref{equilibrio})
and the stability one (\ref{estabilidad}) are invariant under
the constant scaling $a(r)\to\Lambda a(r)$, although the reduced
Einstein equations are not.]

It is easy to check that condition (\ref{equilibrio}) cannot be fulfilled
by a membrane in the Minkowski spacetime (as had been already shown
in Ref.\ \cite{C94}). The case is however completely different when the
background is curved. Consider, for example, a Schwarzschild black hole.
Then $a(r)=b(r)=1-2M/r$ (taking now $G=1$). In this case
Eq.\ (\ref{equilibrio}) leads to
\begin{equation}
{T\over C}={2M\over(2r_m^3-3Mr_m^2)}~,
\end{equation}
which can be fulfilled for any value of $r_m$ (outside the event horizon)
by a suitable choice of $T/C$. This membrane, however, is in an unstable
equilibrium since the inequality (\ref{estabilidad}) is always violated
for $a(r)=b(r)=1-2M/r$. It is however easy to construct by hand
"black hole" metrics $a(r),\;b(r)$ such that
Eqs.\ (\ref{equilibrio})--(\ref{estabilidad}) can be
fulfilled somewhere outside the horizon, see Section VI.

\section{Self-consistent static membrane}

So far we have considered the membrane propagating in an arbitrary
curved background. In this section we will consider the self--consistent
problem of computing the spacetime generated by the membrane itself.
In general this is a very complicated system of non--linear equations.
We shall restrict the study to the spherically symmetric case and where
the only matter field is represented by the membrane of the previous section.

The total action will then consist of the following two pieces
\begin{eqnarray}
S_G&=&{1\over16\pi}\int{}^{(4)}R\sqrt{-g}\;d^4x~,\\
S_m&=&\int d\tau d\rho d\sigma \sqrt{-\gamma}\;(-T+C\;{}^{(3)}R)~.
\label{acciones}
\end{eqnarray}

The usual functional variation with respect to the metric $g_{\mu\nu}$
gives the Einstein equations
\begin{equation}
G_{\mu\nu}\equiv R_{\mu\nu}-{1\over2}Rg_{\mu\nu}=8\pi T_{\mu\nu}~,
\label{ee}
\end{equation}
where the stress--energy--momentum tensor of the membrane is given by
\begin{equation}
\sqrt{-g}\;T^{\mu\nu}=-2\int d\tau d\rho d\sigma \sqrt{-\gamma}\;
\delta\left(x^\lambda-X^\lambda(\tau,\rho,\sigma)\right)
\left\{{T\over2}\gamma^{ij}+C\left({}^{(3)}R^{ij}-{1\over2}
{}^{(3)}R\gamma^{ij}\right)\right\}X^\mu_{,i}X^\nu_{,j},
\label{tsem}
\end{equation}
which for a static spherical membrane at $r=r_m$ takes the following
explicit form (where we have performed the integrals in\ (\ref{tsem}))
\begin{eqnarray}
T_{tt}&=&{a\sqrt{b}\over r^2}(Tr^2-2C)\delta(r-r_m)~,\nonumber\\
T_{rr}&=&0~\nonumber,\\
T_{\theta\theta}&=&-T\sqrt{b}r^2\delta(r-r_m)~,\label{4.5}\\
T_{\phi\phi}&=&\sin^2\theta T_{\theta\theta}~.\nonumber
\end{eqnarray}
Note that there is no radial pressure and that the scalar
curvature term only contributes to the energy density.
The total energy of the membrane is given by
\begin{equation}
\mbox{Energy}=-\int\sqrt{-g}\;T^t\;_{t}d^3\vec{x}=4\pi\sqrt{a(r_m)}\;
(Tr_m^2-2C)=E~,\label{4.6}
\end{equation}
which coincides with the energy defined through the
Hamiltonian\ (\ref{hamiltoniano}), see equation (\ref{equilibrio1}).

The Einstein equations\ (\ref{ee}) are solved\ \cite{LS88} by
\begin{eqnarray}
b(r)&=&1-{2M\over r}-{2E\over r}\Theta(r-r_m)~,\label{4.10}\\
a(r)&=&b(r)\exp\left\{{-E(1-2\Theta(r-r_m))\over r_m-2M-E}\right\}~,
\label{metrica}
\end{eqnarray}
where we have chosen integration constants such that $a(r_m)=b(r_m)$,
and where
\begin{eqnarray}
E&=&4\pi\sqrt{1-{2M\over r_m}-{E\over r_m}}\;(Tr_m^2-2C)~,\label{4.12}\\
-16\pi T&=&{E\over r_m^3}(2M+E)
\left(1-{2M\over r_m}-{E\over r_m}\right)^{-3/2}~,
\label{et}
\end{eqnarray}
and our conventions are such that $\Theta(0)=1/2.$
Note that equations (\ref{4.12})--(\ref{et}) as obtained from the
Einstein equations are exactly equivalent
to equations (\ref{equilibrio1})--(\ref{equilibrio}) as
obtained from the stability conditions, i.e. from the
equations of motion for the membrane. We thus
conclude that the static spherical
membrane at $r_m$ {\em is} a self--consistent solution in the
spacetime (\ref{4.10})--(\ref{metrica}).
The string tension $T$ is positive and we
shall also assume that the integration constant
$M$ is positive. It follows that inside the
spherical membrane, the spacetime is Schwarzschild
corresponding to mass $M$ whi
le outside it is Schwarzschild corresponding to mass
$M+E,$ as expected since $E$ is the energy of the membrane. However, for
Eq.\ (\ref{et}) to be fulfilled, $E$ must be negative (since we
assumed $T$ and $M$ positives). In addition the
inequality\ (\ref{estabilidad}) is not satisfied, and we conclude
that the static self-consistent membrane is in unstable equilibrium.

\section{Membrane with generic second derivative terms}

In this section we shall discuss the possibility of generalizing the
analysis of sections III and IV to the generic case of second order membranes,
as described by the action (\ref{IV.1}).

The effective Lagrangian is obtained from Eqs.\ (\ref{2.9}),
(\ref{2.13}), (\ref{2.18}) and (\ref{2.19}),
and we can then construct the Hamiltonian via the generalized
Legendre transform as in
Eqs.\ (\ref{3.2})--(\ref{3.4}). This Hamiltonian will now depend
on $r,\;\partial_\tau r,\;\partial^2_\tau r$ and $\partial^3_\tau r,$ so we
will not have a simple description of the dynamics in terms of an
effective potential, in
the form of an equation like (\ref{3.6}).
The conserved Hamiltonian ``energy'' is
however still obtained from ${\cal H}= E$ = constant, and the condition for
having an equilibrium configuration can be obtained from the Euler-Lagrange
equation (\ref{3.3'}) by setting
$\partial_\tau r=\partial^2_\tau r=\partial^3_\tau r=\partial^4_\tau r=0.$
The two equations generalizing
(\ref{equilibrio1})--(\ref{equilibrio}) thus become
\begin{equation}
E=4\pi\sqrt{a}\left\{ (Tr^2-2C)-(A+B)\left( 2b+\frac{br^2a'^2}{4a^2}\right)-
A\left(2b+\frac{2rba'}{a}\right)\right\},
\end{equation}
\begin{eqnarray}
T(r^2a'+4ar)&=&2Ca'+A\left( 4b'a+6ba'-\frac{2rba'^2}{a}+4rba''+4ra'b'\right)
\nonumber\\
&+&(A+B)\left(2ba'+4ab'-\frac{3br^2a'^3}{4a^2}+\frac{r^2b'a'^2}{2a}+
\frac{bra'^2}{a}+\frac{br^2a'a''}{a}\right),
\end{eqnarray}
and $r$ has to be evaluated everywhere at the equilibrium position $r_m.$
Again we notice that these equations cannot be fulfilled in Minkowski
space\cite{C94}. They can however be easily fulfilled in most curved
spacetimes, for instance the Schwarzschild spacetime, as was shown by the
present authors in \cite{LL96}.

Next we have to consider the question of stability of the equilibrium
configurations. Since we do not, in this case, have a potential $V(r)$
as in Eq.\ (\ref{3.6}), we proceed as in Ref. \cite{LL96}. 
Introduce the function $\phi(r)$
\begin{equation}
r=r_m+\phi(\tau),
\end{equation}
and expand the Euler-Lagrange equation (\ref{3.3'})
to first order in $\phi.$ After some
algebra, the resulting differential equation determining the radial
fluctuations takes the general form
\begin{equation}
\frac{d^4\phi}{d\tau^4}+F(r_m)\frac{d^2\phi}{d\tau^2}+
G(r_m)\phi=0~,
\label{IV.11}
\end{equation}
where $F(r_m)$ and $G(r_m)$ are complicated functions carrying the
information
about the static zeroth order solution and of the curved spacetime.

In the most general (non-degenerate) case, this fluctuation equation is
solved by
\begin{equation}
\phi(\tau)=c_1 e^{d_1\tau}+c_2 e^{d_2\tau}+c_3 e^{d_3\tau}+
c_4 e^{d_4\tau},
\end{equation}
where $(c_1,c_2,c_3,c_4)$ are arbitrary constants, and
\begin{equation}
d_{(1,2,3,4)}=\pm\left( \frac{-F(r_m)\pm\sqrt{F^2(r_m)-4G(r_m)}}{2}
\right)^{1/2}.
\end{equation}
The necessary and sufficient condition for stability
is that $d_{(1,2,3,4)}$ are all purely imaginary, corresponding to
$\phi(\tau)$ being  oscillatory.

Here we shall not give the (rather complicated) general expressions for
the functions $F(r_m),\:G(r_m).$ They were given in Ref. \cite{LL96} for
the case of the Schwarzschild spacetime (for which the constant $C$ can
be set equal to zero without loss of generality, c.f. Eq.\ (\ref{2.6})),
and were shown to lead to the conclusion that
a static spherical equilibrium membrane in the Schwarzschild spacetime is
always unstable. This is not, however, the case in a general curved
spacetime; see the next section.

We now briefly discuss the  remaining question of the self-consistency of
the equilibrium solutions for the generic second-derivative membranes.
First one has to compute the contributions to the stress-energy-momentum
tensor coming from the terms proportional to $A$ and $B$ in 
the action (\ref{IV.1}). For this purpose, it is convenient to eliminate the
normal-vectors using the completeness-relation and the Gau\ss-Weingarten
equation,
\begin{equation}
(\gamma^{ij}\Omega_{ij})^2=g_{\mu\nu}(\Box x^\mu+\gamma^{ij}
\Gamma^\mu_{\rho\sigma}x^\rho_{,i}x^\sigma_{,j})(\Box x^\nu+\gamma^{ij}
\Gamma^\nu_{\rho\sigma}x^\rho_{,i}x^\sigma_{,j}),
\end{equation}
\begin{equation}
\Omega^{ij}\Omega_{ij}=\gamma^{ik}\gamma^{jl}(g_{\mu\rho}x^
\lambda_{,i}\nabla_\lambda x^\rho_{,j})(g_{\nu\rho}x^\lambda_{,k}
\nabla_\lambda x^\rho_{,l})
(g^{\mu\nu}-\gamma^{mn}x^\mu_{,m}x^\nu_{,n}).
\end{equation}
Now it is straightforward (although tedious) to perform the functional
variation with respect to the metric $g_{\mu\nu}.$ Here we will only give
the results for the static spherical membrane at $r=r_m.$ From the
$(\gamma^{ij}\Omega_{ij})^2$ term we get
\begin{eqnarray}
\Delta_A
T_{tt}&=&2\sqrt{b}\left( \frac{-2ab}{r^2}+\frac{3ba'^2}{8a}+\frac{a'b}{r}
\right)\delta(r-r_m)-2\sqrt{b}\left(\frac{2ab}{r}+\frac{a'b}{2}\right)
\delta'(r-r_m)~,\\
\Delta_A
T_{rr}&=&-2\sqrt{b}\left( \frac{4}{r^2}+\frac{a'^2}{4a^2}+\frac{2a'}{ar}
\right)\delta(r-r_m),~\\
\Delta_A
T_{\theta\theta}&=&2\sqrt{b}\left(-2b+\frac{br^2a'^2}
{8a^2}\right)\delta(r-r_m)+2\sqrt{b}\left(2br+\frac{r^2a'b}{2a}\right)
\delta'(r-r_m)~,\\
\Delta_A
T_{\phi\phi}&=&\sin^2\theta\; T_{\theta\theta}.
\end{eqnarray}
The $\Omega^{ij}\Omega_{ij}$ term gives rise to
\begin{eqnarray}
\Delta_B
T_{tt}&=&2\sqrt{b}\left( \frac{-ab}{r^2}+\frac{3ba'^2}{8a}\right)
\delta(r-r_m)-\sqrt{b} a'b\; \delta'(r-r_m)~,\\
\Delta_B
T_{rr}&=&-2\sqrt{b}\left( \frac{2}{r^2}+\frac{a'^2}{4a^2}\right)
\delta(r-r_m),~\\
\Delta_B
T_{\theta\theta}&=&2\sqrt{b}\left(-b+\frac{br^2a'^2}
{8a^2}\right)\delta(r-r_m)+2\sqrt{b} br\; \delta'(r-r_m)~,\\
\Delta_B
T_{\phi\phi}&=&\sin^2\theta \;T_{\theta\theta}.
\end{eqnarray}
Note that the expressions for $T_{tt}$ and $T_{rr}$ can also (and more
easily) be obtained directly from the effective Lagrangian by taking the
functional variations with respect to $-a(r)$ and $1/b(r),$ respectively.

Together with equations (\ref{4.5}), we now have the complete expressions
for the stress-energy-momentum tensor for a static spherical membrane
described by the action (\ref{IV.1}).
However, the presence of the $\delta'(r-r_m)$-terms is
somewhat problematic. It is still possible to define the
total energy of the membrane along the lines of (\ref{4.6}), and
to get a well-defined result. The $(tt)$ and $
(rr)$ components of the
Einstein equations (\ref{ee}) can also be easily integrated\cite{LS88},
at least formally. However, due to the
$\delta'(r-r_m)$-terms in $T_{tt},$ we obtain ill-defined expressions
for the functions $a(r)$ and $b(r)$ involving products of delta-functions
and exponentials of delta-functions. Presently it is thus not clear
how to interpret the results physically, at least not without some
kind of  regularisation of the singular functions.
We therefore leave this problem for further study elsewhere.

\section{Discussion}

We have extended the analysis of the stability of spherical
membranes in curved spacetimes, in terms of an effective potential,
to the case of higher order membranes with a Lagrangian dependence
proportional to ${}^{(3)}R$. We have also
considered the problem of stability
for the higher order membrane Lagrangian including all possible second
derivative terms. We have found static solutions in fixed background
spacetimes, and we have also been able
to find self--consistent static solutions. However, these
solutions are unstable against small radial perturbations. This
does not mean, however,  that we are ready
to infer that {\em no} membranes
can be kept in stable equilibrium outside a black hole. For that,
one should prove a ``no--hair'' theorem for membranes including also
other types of matter on the right hand side of the Einstein equations.
At this point it is not clear whether it is possible or not to prove
such theorem.

On the
other hand it is relatively easy to construct,
by hand, arbitrary functions
$a(r)$ and $b(r)$ describing a black hole (plus a source with a
positive mass density contribution) such that the stability 
Eqs.\ (\ref{equilibrio})--(\ref{estabilidad})
are satisfied for some $r_m$ outside
the horizon. As an example, consider
\begin{equation}
a(r)=b(r)={2\over3}-{2M\over3r}+{1\over3}
\tanh\left({16r\over M}-40\right)~,
\end{equation}
for some $r<R$ and then smoothly matched to the Schwarzschild
metric. This functional form of the line--element allows for
static {\it stable} membranes just outside the event horizon as can
be easily verified. Such a spacetime can however not be supported by
only the
membrane itself, as follows from our analysis, but at this point it is
yet not clear whether inclusion of some other kinds of "normal" matter
could help on that.

\begin{acknowledgments}
A.L.L. was supported by NSERC (National
Sciences and Engineering Research Council of Canada), while
C.O.L was supported by NSF grant No. PHY-95-07719 and by research
funds of the University of Utah.
\end{acknowledgments}

\end{document}